\documentclass[titlepage,letterpaper,12pt]{article}
\begin{document}

\title{Comment on \\
``Kepler problem in Dirac theory for a particle with position-dependent
mass''}
\date{}
\author{Antonio S. de Castro \\
%EndAName
\\
UNESP - Campus de Guaratinguet\'{a}\\
Departamento de F\'{\i}sica e Qu\'{\i}mica\\
Caixa Postal 205\\
12516-410 Guaratinguet\'{a} SP - Brasil\\
\\
E-mail address: castro@feg.unesp.br (A.S. de Castro)}
\date{}
\maketitle

In a recent paper, Vakarchuk \cite{vak} approached the Dirac equation
coupled to a vector Coulomb potential for a particle with an effective mass
dependent on the position. He assumed that the effective mass has a form of
a multipole expansion and took only the first two lowest terms into account.
In effect, he considered a fermion in the background of a mixed
vector-scalar Coulomb potential.

Back in 1973 Soff \cite{sof} et al. found the analytic solution of the Dirac
equation with an arbitrary mixing of vector and scalar potentials. The
solution for the spin and angular variables was expressed in terms of spinor
spherical harmonics, also called spherical spinors, resulting from the
coupling among two-component spinors and spherical harmonic functions. The
radial equations for the upper and lower components of the Dirac spinor were
treated by the brute force power series expansion method. In a recent time
the solution of this problem was used to speculate about the breaking of
pseudospin symmetry in heavy nuclei \cite{gin}. It is also worthwhile to
mention that a pedagogical approach to the Dirac equation coupled to a mixed
vector-scalar Coulomb potential is already crystallized in a textbook \cite
{gre}.

It seems that Vakarchuk did not know about the solution of the Dirac
equation with a mixed vector-scalar Coulomb potential and, as a matter of
fact, he presented a more elegant method to find out the analytic solution.
Although it may seem strange, the author tried to map the Dirac equation
into an effective Schr\"{o}dinger equation for all the components of the
Dirac spinor. Nevertheless, I am afraid that something might be by far wrong.

In order to clarify my criticism, let me begin writing the Dirac equation
with the effective mass and the Coulomb interaction as given in \cite{vak}:
\begin{equation}
\left( c\mathbf{\hat{\alpha}.\hat{p}}+m^{*}c^{2}\hat{\beta}+U\right) \psi
=E\psi ,\quad m^{*}=m\left( 1+a/r\right) ,\quad U=-e^{2}/r  \label{1}
\end{equation}

\noindent where $a$ and $e$ are constants. With $\bar{\psi}$ defined as
\begin{equation}
\psi =\left( c\mathbf{\hat{\alpha}.\hat{p}}+m^{*}c^{2}\hat{\beta}+E-U\right)
\bar{\psi}  \label{2}
\end{equation}

\noindent the author of Ref. \cite{vak} obtains
\[
\left[ \frac{\mathbf{\hat{p}}^{2}}{2m}-\frac{e^{2}E/\left( mc^{2}\right)
-mc^{2}a}{r}+\frac{\left( i\hbar /c\right) \mathbf{\sigma .n}\left( mc^{2}a%
\hat{\beta}^{\prime \prime }+e^{2}\hat{\beta}^{\prime }\right)
+m^{2}c^{2}a^{2}-e^{4}/c^{2}}{2mr^{2}}\right] \bar{\psi}
\]
\begin{equation}
\hfill \hspace{\stretch{10}}=\frac{E^{2}-m^{2}c^{4}}{2mc^{2}}\,\bar{\psi}
\label{3}
\end{equation}

\noindent where $\mathbf{n}=\mathbf{r}/r$, $\mathbf{\sigma }$ stands for the
Pauli matrices and the 4$\times $4 matrices $\hat{\beta}^{\prime }$ and $%
\hat{\beta}^{\prime \prime }$ are defined as
\begin{equation}
\hat{\beta}^{\prime }=\left(
\begin{array}{ll}
0 & I \\
I & 0
\end{array}
\right) ,\quad \hat{\beta}^{\prime \prime }=\left(
\begin{array}{ll}
0 & -I \\
I & 0
\end{array}
\right)  \label{4}
\end{equation}

\noindent Instead of $\mathbf{\sigma }$, a 4$\times $4 matrix whose block
diagonal elements are Pauli matrices, $\mathbf{\sigma }$, and whose
off-diagonal block elements are zero, should used. Furthermore, contrary to
the statement of author of Ref. \cite{vak}, Eq. (\ref{3}) does not have the
form of a Schr\"{o}dinger equation. This criticism is endorsed by observing
that the ``centrifugal barrier'' term contains off-diagonal matrix elements
which mix the upper and lower components of the quadrispinor $\bar{\psi}$.

By introducing the operator
\begin{equation}
\hat{\Lambda}=-\left( \mathbf{\sigma .\hat{L}+}\hbar \right) +\frac{i}{c}%
\mathbf{\sigma .n}\left( mc^{2}a\hat{\beta}^{\prime \prime }+e^{2}\hat{\beta}%
^{\prime }\right)   \label{5}
\end{equation}
\noindent where $\mathbf{\hat{L}}$ is the angular momentum operator, Eq. (%
\ref{3}) in the spherical coordinate system is supposed to reduce to \cite
{vak}:

\begin{equation}
\left[ -\frac{\hbar ^{2}}{2m}\frac{1}{r}\frac{d^{2}}{dr^{2}}\,r-\frac{%
e^{2}E/\left( mc^{2}\right) -mc^{2}a}{r}+\frac{\hbar ^{2}l^{*}\left( l^{*}%
\mathbf{+}1\right) }{2mr^{2}}\right] R=\frac{E^{2}-m^{2}c^{4}}{2mc^{2}}\,R
\label{6}
\end{equation}

\noindent where $R$ is the radial part of $\bar{\psi}$, $\hbar
^{2}l^{*}\left( l^{*}\mathbf{+}1\right) $ is the eigenvalue of the operator $%
\hat{\Lambda}\left( \hat{\Lambda}\mathbf{+}\hbar \right) $ for some
eigenfunction, of course, and
\begin{equation}
l^{*}=\sqrt{\left( j+\frac{1}{2}\right) ^{2}+\left( \frac{mca}{\hbar }%
\right) ^{2}-\left( \frac{e^{2}}{\hbar c}\right) ^{2}}-\frac{1}{2}\mp \frac{1%
}{2}  \label{7}
\end{equation}

\noindent with the upper sign for $j=l+1/2$ and the lower sign for $j=l-1/2$%
. Note that $\hat{\Lambda}$ operates on the spin and angular variables. It
is a generalization of the spin-orbit coupling operator $\hat{K}$, part of
the toolkit for studying the Dirac equation with spherically symmetric
potentials (see, e.g., \cite{sof}-\cite{gre}), and has the spinor spherical
harmonic as eigenfunction. The eigenfunction of the operator $\hat{\Lambda}$
is not specified in Ref. \cite{vak}. In that case, what is the generalized
spinor spherical harmonic?

The author of Ref. \cite{vak} affirms that Eq. (\ref{6}) formally coincides
with the nonrelativistic Schr\"{o}dinger equation for the Kepler problem.
This declaration would be faithful if $l^{*}$ could only assume nonnegative
integer values but it does not in general. Therefore, one can neither
identify $R$ with the usual radial functions of the nonrelativistic hydrogen
problem $R_{nl}$, with $n=0,1,2,\ldots $ and $l=0,1,2,\ldots ,(n-1)$ nor
identify $\left( E^{2}-m^{2}c^{4}\right) /(2mc^{2})$ with the ``Bohr
formula'' for the energy levels. Notice, though, that the identification is
believed to be true when $e^{2}=mc^{2}|a|$, i.e., when the scalar potential,
if it is either attractive or repulsive, is as strong as the vector
potential.

If one insists that (\ref{6}) is the nonrelativistic Schr\"{o}dinger
equation for the Kepler problem then the Dirac eigenenergy solution (Eq.
(4.2) in Ref. \cite{vak}) should have two branches of solutions,
corresponding to positive and negative energy in a general circumstance.
Furthermore, the condition to the existence of bounded solutions should be
written as $a<e^{2}E/\left( m^{2}c^{4}\right) $ and not $a<e^{2}/\left(
mc^{2}\right) $.

Based on the above considerations it is not difficult to be vehemently
opposed not only the solutions found in Ref. \cite{vak} but also the
conclusions manifested there.

\vspace{1in}

\noindent This work was supported in part by means of funds provided by CNPq
and FAPESP.

\end{document}